\documentstyle[aps,amssymb,12pt,psfig]{revtex}
\begin{document}
\tightenlines
\title{Optimized quantum nondemolition measurement of a field quadrature} 
\author{Matteo G. A. Paris}\address{Quantum Optics $\&$ Information Group, 
Istituto Nazionale per la Fisica della Materia \\ 
Universit\`a di Pavia, via Bassi 6, I-27100 Pavia, Italy}
\maketitle
\begin{abstract}
We suggest an interferometric scheme assisted by squeezing and linear feedback
to realize the whole class of field-quadrature quantum nondemolition
measurements, from Von Neumann projective measurement to fully non-destructive
non-informative one.  In our setup, the signal under investigation is mixed
with a squeezed probe in an interferometer and, at the output, one of the two
modes is revealed through homodyne detection. The second beam is then
amplitude-modulated according to the outcome of the measurement, and finally
squeezed according to the transmittivity of the interferometer.  Using
strongly squeezed or anti-squeezed probes respectively, one achieves either a
projective measurement, {\em i.e.} homodyne statistics arbitrarily close to
the intrinsic quadrature distribution of the signal, and conditional outputs
approaching the corresponding eigenstates, or fully non-destructive one,
characterized by an almost uniform homodyne statistics, and by an output state
arbitrarily close to the input signal. By varying the squeezing between these
two extremes, or simply by tuning the internal phase-shift of the
interferometer, the whole set of intermediate cases can also be obtained. In
particular, an optimal quantum nondemolition measurement of quadrature can be 
achieved, which minimizes the information gain versus state disturbance trade-off.  
\end{abstract}
\section{Introduction}
In order to be manipulated and transmitted information should be encoded into
some degree of freedom of a physical system. Ultimately, this means that the
input alphabet should correspond to the spectrum of some observable, {\em
i.e.} that information is transmitted using {\em quantum signals}. At the end
of the channel, to retrieve this kind of quantum information, one should
measure the corresponding observable. As a matter of fact, the measurement
process unavoidably introduces some disturbance, and may even destroys the
signal, as it happens in many quantum optical detectors, which are mostly
based on the irreversible absorption of the measured radiation. Actually, even
in a measurement scheme that somehow preserves the signal for further uses,
one is faced by the information gain versus state disturbance trade-off, {\em
i.e.} by the fact that the more information is obtained, the more the signal
under investigation is being modified. 
\par
Actually, the most informative measurement of an observable $X$ on a state
$|\psi\rangle$ corresponds to its ideal projective measurement, which is also 
referred to as Von Neumann {\em second kind} quantum measurement \cite{vn}. 
In an ideal projective measurement the outcome $x$ occurs with the
intrinsic probability density $|\langle\psi|x\rangle|^2$, whereas the system
after the measurement is left in the corresponding eigenstate $|x\rangle$.  A
projective measurement is obviously repeatable, since a second measure gives
the same outcome as the first one. However, the initial state is erased, and 
the conditional output do not permit to obtain further information about the 
input signal.  
The opposite case corresponds to a fully non-destructive detection scheme,
where the state after the measurement can be made arbitrarily close to the
input signal, and which is characterized by an almost uniform output
statistics, {\em i.e.} by a data sample that provides almost no information.
\par
Besides fundamental interest, the realization of a projective measurement of the
quadrature would have application in quantum communication based on continuous
variables. In facts, it provides a reliable and controlled source of optical
signals. On the other hand, a fully non-destructive measurement scheme is an
example of a quantum repeater, another relevant tool for the realization of
quantum network. 
Between these two extremes we have the entire class of quantum nondemolition
(QND) measurements. Such intermediate schemes provide only a partial information
about the measured observable, and correspondingly are only partially
distorting the signal under investigation. In particular, in this 
paper, we show how to attain an optimized QND measurement of quadrature 
{\em i.e.} scheme which minimizes the information gain versus state disturbance trade-off.
\par
Most of the schemes suggested for back-action evading measurements are based
on nonlinear interaction between signal and probe taking place either in
$\chi^{(3)}$ or $\chi^{(3)}$ media (both fibers and crystals)
\cite{wals,lap,per,gra,yam,bruck,haus}, or on optomechanical coupling
\cite{opt1,opt2}. A beam-splitter based scheme has been earlier suggested to
realize optical Von Neumann measurement \cite{vnm}. Here, we focus our
attention on an interferometric scheme which requires only linear elements and
single-mode squeezers. \par
\begin{figure}[h]
\psfig{file=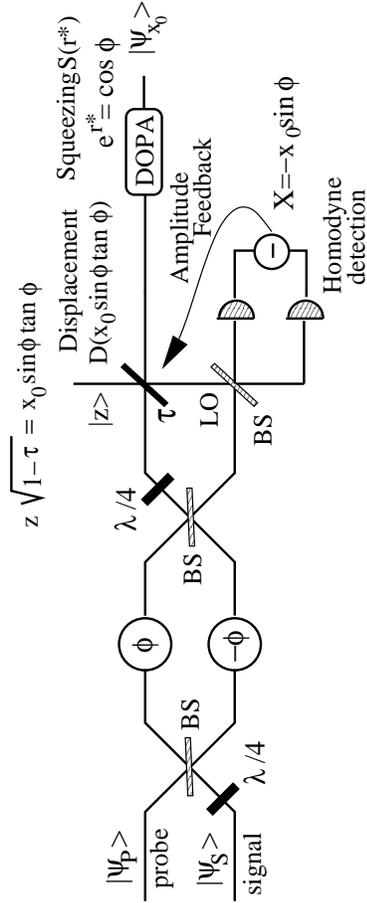,width=10cm}
\caption{Setup for QND measurements of a field
quadrature on the state $|\psi_{\sc s}\rangle$. The signal is
linearly mixed with a probe mode in a Mach-Zehnder interferometer,
which, equipped with two additional $\lambda/4$ plates, shows a 
transmittivity $\cos^2\phi$.  Such
transmittivity can be tuned by varying the internal phase-shift
$\phi$.  One of mode exiting the interferometer is then revealed by
homodyne detection, and the resulting outcome $x_0$ is used for a
feedback-assisted displacement, by an amount $x_0\sin\phi\tan\phi$ of
the other mode. Such displacement is obtained by mixing the mode with
a strong coherent state of amplitude $z$ ({\em e.g.} the laser beam
also used as local oscillator for the homodyne detector) in beam
splitter of transmittivity $\tau$ close to unit, with the requirement
that $z\sqrt{1-\tau}=x_0\sin \phi\tan\phi$. Finally, the conditional
output is squeezed by a degenerate parametric amplifier (DOPA) by an
amount $S(r^\star)$ with $e^{r^\star}=\cos\phi$. By varying the degree 
of squeezing of the probe mode the resulting measurement ranges 
from a projective to a fully non-destructive detection of the field 
quadrature.}\label{f:setup}
\end{figure}

A schematic diagram of the suggested setup is given in Fig. \ref{f:setup}.
The signal under examination $|\psi_{\sc s}\rangle$ and the probe (meter) 
state $|\psi_{\sc p}\rangle$ are given by
\begin{eqnarray}
|\psi_{\sc s}\rangle &=& \int dx \: \psi_{\sc s} (x) \:
|x\rangle_1 \nonumber \\ |\psi_{\sc p}\rangle &=& \int dx \:
\psi_{\sc p} (x) \: |x\rangle_2 \label{defstate}\;,
\end{eqnarray}
where $|x\rangle_j$, $j=1,2$ are eigenstates of the field quadratures
$x_j=1/2(a_j^\dag+a_j)$, $j=1,2$ of the two modes, and $\psi_{\sc s}(x)$ and
$\psi_{\sc p}(x)$ are the corresponding wave-functions. The two beams are
linearly mixed in a Mach-Zehnder interferometer with internal phase-shift
given by $\phi$. There are also two $\lambda/4$ plates, each imposing a
$\pi/2$ phase-shift.  Overall, the interferometer equipped with the plates is
equivalent to a beam splitter of transmittivity $\tau=\cos^2\phi$.  However,
the interferometric setup is preferable to a single beam splitter since it
permits a fine tuning of the transmittivity. After the interferometer, one of
the two output modes is revealed by homodyne detection, whereas the second
mode is firstly displaced by an amount that depends on the outcome of the
measurement (feedback assisted amplitude modulation), and then squeezed
according to the transmittivity of the interferometer (see details below).  As
we will see, either by tuning the phase-shift of the interferometer, or by
exciting the probe state $|\psi_{\sc p}\rangle$ in a squeezed vacuum, and by 
varying the degree of squeezing, the action of the setup ranges from a 
projective to a non-destructive measurement of the field quadrature as 
follows:
\begin{enumerate}
\item the statistics of the homodyne detector ranges from a distribution 
arbitrarily close to the intrinsic quadrature probability density of the 
signal state $|\psi_{\sc s} (x)|^2$ to an almost uniform distribution; 
\item the conditional output state, after registering a value $x_0$ for the 
quadrature of the signal mode, ranges from a state arbitrarily close to
the corresponding quadrature eigenstate $|x_0\rangle$ 
to a state that approaches the input signal $|\psi_{\sc s}\rangle$.
\end{enumerate}
The two features can be summarized by saying that the present scheme
realizes the whole set of QND measurements of a field quadrature.
In addition, the interferometer can be tuned in order to
minimize the information gain versus state disturbance trade-off, 
{\em i.e.} to achieve an optimal QND measurement of quadrature.
Such a kind of measurement provides the maximum information about
the quadrature distribution of the signal, while keeping the
conditional output state as close as possible to the incoming signal. 
\par
The paper is structured as follows. In the next Section we analyze the
dynamics of the measurement scheme, and describe in details the action of
linear feedback and tunable squeezing on the conditional output state and on
the homodyne distribution. In Section \ref{s:lim} we analyze the limiting
cases of strongly squeezed and anti-squeezed probes, which correspond to
projective and non-destructive  measurements respectively. In Section
\ref{s:opt} we introduce two fidelity measures, in order to quantify how close
are the conditional output and the homodyne distribution to the input signal
and its quadrature distribution respectively. As a consequence, we are able to
individuate an optimal set of configurations that minimize the trade-off
between information gain and  state disturbance. Section \ref{s:outro} closes
the paper with some concluding remarks.
\section{Homodyne interferometry with linear feedback}
Let us now describe the interaction scheme in details. The evolution
operator of the interferometer is given by $U(\phi) = \exp
\left[i\phi \left(a_1^\dag a_2 + a^\dag_2 a_1\right)\right]$, 
such that the input state $|\Psi_{\sc in}\rangle\rangle =
|\psi_{\sc s}\rangle \otimes |\psi_{\sc p}\rangle$ evolves as  
\begin{eqnarray}
|\Psi_{\sc out}\rangle\rangle &=& U(\phi)\: |\psi_{\sc s}\rangle \otimes 
|\psi_{\sc p}\rangle = \nonumber \\&=& \int dx_1\int dx_2 \:
\psi_{\sc s} (x_1)\:\psi_{\sc p} (x_2)\: |x_1\cos\phi+x_2\sin\phi\rangle_1
\otimes|-x_1\sin\phi+x_2\cos\phi\rangle_2 \nonumber \\&=&
\int dy_1\int dy_2 \: \psi_{\sc s} (y_1\cos\phi-y_2\sin\phi )
\: \psi_{\sc p}(y_1\cos\phi+y_2\sin\phi)\: |y_1\rangle_1
\otimes|y_2\rangle_2
\label{outMZ}\;.
\end{eqnarray}
After the interferometer the quadrature of one of the modes 
(say mode $2$) is revealed by homodyne detection. The distribution 
of the outcomes is given by 
\begin{eqnarray}
p(X)=\hbox{Tr}\left[\: |\Psi_{\sc
out}\rangle\rangle\langle\langle\Psi_{\sc out}|\: {\mathbb I}_1\otimes
\Pi_2(X)\right] \qquad \Pi(X)=|X\rangle\langle X|
\label{probX}\;,
\end{eqnarray}
$\Pi(x)$ being the POVM of the homodyne detector.
Since the reflectivity of the interferometer is given by $\sin\phi$ 
from an outcome $X$ by the homodyne we infer a value $x_0=-X/\sin\phi$ 
for the quadrature of the input signal. The corresponding probability  
density is given by \begin{eqnarray}
p(x_0)=-\sin\phi \: p(X) = \tan\phi \int dy\: |\psi_{\sc s}(y)|^2 \: 
\left|\psi_{\sc p}\left[\tan\phi(y-x_0)\right] \right|^2
\label{probx0}\;,
\end{eqnarray}
and the conditional output state for the mode $1$
\begin{eqnarray}
|\varphi_{x_0}\rangle&=& \sqrt{\sin\phi}\: 
\langle -X\sin\phi|\Psi_{\sc out}\rangle\rangle =
\nonumber \\ 
&=& \sqrt{\frac{\sin\phi}{p(x_0)}} \int dy \:
\psi_{\sc s}(y\cos\phi+x_0\sin^2\phi)\:
\psi_{\sc p}(y\sin\phi-x_0\cos\phi\sin\phi)\:|y\rangle
\label{cond1}\;.
\end{eqnarray}
The amplitude of this conditional state is then modulated by a 
feedback mechanism, which consists in the application of a 
displacement $D(x_0\sin\phi\tan\phi)$, with $D(z)=\exp(za^
\dag-\bar{z}a)$. 
Such displacing action can be obtained by mixing the mode 
with a strong coherent state of amplitude $z$ ({\em e.g.} the laser beam 
also used as local oscillator for the homodyne detector, see 
Fig. \ref{f:setup}) in a beam splitter of transmittivity $\tau$ close 
to unit, with the requirement that $z\sqrt{1-\tau}=x_0\sin\phi\tan\phi$ 
\cite{displa}. An experimental implementation using a feedforward
electro-optic modulator has been presented in \cite{lam}.
The resulting state is given by 
\begin{eqnarray}
D(x_0\sin\phi\tan\phi)\:|\varphi_{x_0}\rangle
= \sqrt{\frac{\sin\phi}{p(x_0)}} \int dy \:
\psi_{\sc s}(y\cos\phi)\:
\psi_{\sc p}(y\sin\phi-x_0\tan\phi)\:|y\rangle
\label{displa}\;.
\end{eqnarray}
Finally, this state is subjected to a single-mode squeezing transformation $S(r)=
\exp[1/2 r (a^{\dag 2}-a^2)]$ by a degenerate parametric amplifier (DOPA).
By tuning the squeezing parameter to a value $r^\star=\cos\phi$ and
using the relation $S(r)|y\rangle=e^{r/2}|e^r y\rangle$ we arrive 
at the final state
\begin{eqnarray}
|\psi_{x_0}\rangle &=& 
S(r^\star)\:D(x_0\sin\phi\tan\phi)\:|\varphi_{x_0}\rangle= \nonumber \\
&=& \sqrt{\frac{\tan\phi}{p(x_0)}} \int dy \:
\psi_{\sc s}(y)\:
\psi_{\sc p}\left[(y-x_0)\tan\phi\right]\:|y\rangle
\label{output}\;.
\end{eqnarray}
The wave-function of this conditional output state is thus given by 
\begin{eqnarray}
\psi_{x_0}(x) =\frac{\psi_{\sc s}(x)\:\psi_{\sc p}\left[(x-x_0)\tan\phi
\right]}{\sqrt{\int dy\: |\psi_{\sc s}(y)|^2 \: 
\left|\psi_{\sc p}\left[\tan\phi(y-x_0)\right] \right|^2}}
\label{outwave}\;.
\end{eqnarray}
Eq. (\ref{probx0}) and Eqs. (\ref{output},\ref{outwave}) summarize the 
filtering effects of the probe wave-function on the output statistics 
and the conditional state respectively. 
\par
\section{Measurements using squeezed or anti-squeezed probes}\label{s:lim}
For the probe mode in the vacuum state we have $\psi_{\sc
p}(x)=(2/\pi)^{-1/4} \exp(-x^2)$ such that the homodyne distribution
of Eq. (\ref{probx0}) results
\begin{eqnarray}
p(x_0)=|\psi_{\sc s}(x)|^2 \star G(x,x_0,\frac{1}{4\tan^2\phi}) 
\label{vacprobx0}\;
\end{eqnarray}
where $\star$ denotes convolution and $G(x;x_0,\sigma^2)$ a Gaussian of 
mean $x_0$ and variance $\sigma^2$. The quadrature distribution of the 
corresponding output state is given by
\begin{eqnarray}
|\psi_{x_0}(x)|^2=\frac{1}{p(x_0)}|\: \psi_{\sc s}(x)|^2\:G(x,x_0,
\frac{1}{4\tan^2\phi})\label{vacond}\;
\end{eqnarray}
Eqs (\ref{vacprobx0}) and (\ref{vacond}) account for the noise introduced 
by vacuum  fluctuations. This noise can be manipulated by suitably squeezing 
the probe, thus realizing the whole set of QND measurement.
\par
Squeezed or anti-squeezed vacuum probes are described by the 
wave-functions
\begin{eqnarray}
\psi_{\sc sq}(x)&=&\frac{1}{(2\pi\Sigma^2)^{1/4}}\exp
\left\{-\frac{x^2}{4\Sigma^2}\right\} \nonumber  \\
\psi_{\sc asq}(x)&=&\left(\frac{\Sigma^2}{2\pi}\right)^{1/4}\exp
\left\{-\frac{\Sigma^2 x^2}{4}\right\}
\label{sqvac}\;, 
\end{eqnarray}
where the information about squeezing stays in the requirement
$0<\Sigma^2\leq 1/4$. Notice that squeezing the probe introduces additional
energy in the system. The mean photon number of the states in 
(\ref{sqvac}) is given by $N=(\Sigma^2 + 1/\Sigma^2 - 2)/4$.
Using a squeezed vacuum probe Eqs. (\ref{vacprobx0}) and (\ref{vacond}) 
rewrites as
\begin{eqnarray}
p(x_0)&=&|\psi_{\sc s}(x)|^2 \star G(x,x_0,\Sigma^2/\tan^2\phi)
\stackrel{\Sigma\rightarrow 0}{\longrightarrow} |\psi_{\sc s}(x_0)|^2 
\label{sqcond1}\\
|\psi_{x_0}(x)|^2&=&\frac{1}{p(x_0)}\:|\psi_{\sc s}(x)|^2\:G(x,x_0,\Sigma^2/\tan^2\phi)
\stackrel{\Sigma\rightarrow 0}{\longrightarrow} \delta(x-x_0)
\label{sqcond2}\;.
\end{eqnarray}
Eq. (\ref{sqcond1}) says that by squeezing the probe the statistics of
the homodyne detectors can be made arbitrarily close to the intrinsic
quadrature distribution $|\psi_{\sc s}(x_0)|^2$, whereas Eq.
(\ref{sqcond2}) shows that, for any value of the outcome $x_0$, the
conditional output $|\psi_{x_0}\rangle$ approaches the corresponding
quadrature eigenstate $|x_0\rangle$.
For $\Sigma\rightarrow 0$ the mean energy of the conditional output
state $|\psi_{x_0}\rangle$ increases, since it is approaching a
quadrature eigenstate (an exact eigenstate would have infinite
energy). Notice that this amount of energy is mostly provided by the
probe state itself, rather than by the displacement and squeezing
stages of the setup. The improvement in the precision due to squeezing, 
compared to that of a vacuum probe, can be quantified by the 
ratio of variances in the filtering Gaussian of Eqs. (\ref{vacond})
and (\ref{sqcond2}). Calling this ratio $\Delta$ we have $\Delta=\Sigma^2$ 
and thus, for squeezing not too low, $\Delta\simeq N$.
\par
For an anti-squeezed vacuum probe Eqs. (\ref{vacprobx0}) and (\ref{vacond}) 
rewrites as
\begin{eqnarray}
p(x_0)&=&|\psi_{\sc s}(x)|^2 \star G(x,x_0,(\Sigma^2 \tan^2\phi)^{-1})
\stackrel{\Sigma\rightarrow 0}{\longrightarrow}
\frac{\exp\{-\frac{x^2}{2\sigma^2}\}}{\sqrt{2\pi\sigma^2}} \quad \sigma^2 =
\frac1{\Sigma^2\tan^2\phi}
\label{sqcond3}\\
|\psi_{x_0}(x)|^2&=&\frac{1}{p(x_0)}\: |\psi_{\sc s}(x)|^2\:G(x,x_0,(\Sigma^2 \tan^2
\phi)^{-1}) \stackrel{\Sigma\rightarrow 0}{\longrightarrow} |\psi_{\sc s}(x)|^2 \quad
\forall x_0
\label{sqcond4}\;.
\end{eqnarray}
Eqs. (\ref{sqcond3}) and (\ref{sqcond4}) says that by anti-squeezing the probe
the statistics of the homodyne detectors is approaching a flat distribution
over the real axis, and correspondingly that the conditional output can be made 
arbitrarily close to the incoming signal, independently on the actual value of 
$x_0$.
\par
Notice that, in principle, both projective and non-destructive measurements
could be obtained with vacuum probe, simply by varying the internal phase-shift of
the interferometer according to Eqs. (\ref{vacprobx0}) and (\ref{vacond}).
However, this would affect also the {\em rate} of the events at the output
(since $\phi$ governs the transmittivity of the interferometer), and
therefore may be not convenient from practical point of view. On the other
hand, when a fine tuning of the variances in Eqs. (\ref{sqcond1}-\ref{sqcond4}) is
needed (as for example in the optimization of the scheme, see next Section) it
can be conveniently obtained by varying $\phi$, without the need of varying the
degree of squeezing of the probe.
\section{Optimized QND measurement}\label{s:opt}
So far we considered the two extreme cases of infinitely 
squeezed or antisqueezed probes. Now we proceed to quanti-fy the trade-off 
between the state disturbance and the gain of information for 
the whole set of intermediate cases. There are two relevant parameters:
i) how close the output signal is to the input state, and ii) how 
close the homodyne distribution is to the intrinsic quadrature probability.
According to Eq. (\ref{output}), after the outcome $x$ is being registered the conditional 
output state is given by $|\psi_x\rangle$. Since the outcome $x$ occurs with
the probability $p(x)$ of Eq. (\ref{probx0}), the density matrix describing the
output ensemble after a large number of measurements is given by 
\begin{eqnarray}
\varrho_{\sc out} = \int dx \: p(x) \: |\psi_x\rangle\langle\psi_x |
\label{outdens}\;.
\end{eqnarray}
Indeed, this is the state that can be subsequently manipulated, or used to 
gain further information on the system. The resemblance between input and 
output can be quantified by the {\em average state fidelity} 
\begin{eqnarray}
F=\langle\psi_{\sc s} | \varrho_{\sc out} | \psi_{\sc s} \rangle = \int dx \:
p(x)\: \left| \langle\psi_{\sc s} | \psi_x\rangle \right|^2
\label{avF}\;.
\end{eqnarray}
Inserting Eq. (\ref{output}) in Eq. (\ref{avF}) we obtain 
\begin{eqnarray}
F=\int\int dy' dy'' \: \left|\psi_{\sc s} (y')\right|^2\:
\left|\psi_{\sc p} (y'')\right|^2\: T_\phi (y',y'')
\label{avF1}\;,
\end{eqnarray}
where, for the squeezed vacuum probes we are taking into account, the transfer function
is given by 
\begin{eqnarray}
T_\phi(y',y'') = \exp \left\{ -\tan^2\phi\:\frac{(y'-y'')^2}{8 \sigma^2_{\sc p}}\right\}
\label{funT}\;,
\end{eqnarray}
$\sigma_{\sc p}^2$ being the variance of the probe wave-function, {\em i.e.}
$\sigma_{\sc p}=\Sigma$ for squeezed probe and $\sigma_{\sc p}=\Sigma^{-1}$
for anti-squeezed probes. F take values from zero to unit, and it is a 
decreasing function of the probe squeezing. 
\par
If also the initial signal is Gaussian the fidelity results
\begin{eqnarray}
F=\frac{\sqrt{2}x}{\sqrt{1+2x^2}}
\qquad x=\frac{\sigma_{\sc p}}{\sigma_{\sc s}\tan\phi}
\label{GF}\;,
\end{eqnarray}
$\sigma_{\sc s}^2$ being the variance of the signal' wave-function.
Eq. (\ref{GF}) interpolates between the two extreme cases of the previous
Section. In order to check this behavior we evaluate (\ref{GF}) for strong
squeezing or anti-squeezing. We have
\begin{eqnarray}
F=\left\{
\begin{array}{lrrc}
\sqrt{2} x     \rightarrow 0 & & x \ll 1 & {\rm squeezed\: probe}\\ & & \\ 
1-(4x^2)^{-1}
 \rightarrow 1 & & x \gg 1 & {\rm antisqueezed\: probe}
\end{array}
\right.
\label{checkF}\;.
\end{eqnarray}
\par
In order to quantify how close the quadrature probability of the
input signal is to the homodyne distribution at the output we introduce the {\em
average distribution fidelity}
\begin{eqnarray}
G= \left(\int dx \: \sqrt{p(x)}\: \left|\psi_{\sc s}(x)\right| \right)^2
\label{G}\;,
\end{eqnarray}
which also ranges from zero to one and it is an increasing function of the
probe squeezing. For both Gaussian signal and probe we obtain
\begin{eqnarray}
G=2\:\frac{\sqrt{1+x^2}}{2+x^2}\qquad x=\frac{\sigma_{\sc p}}{\sigma_{\sc s}\tan\phi}
\label{GG}\;,
\end{eqnarray}
and therefore
\begin{eqnarray}
G=\left\{
\begin{array}{lrrc}
1-\frac18 x^4  \rightarrow 1 & & x \ll 1 & {\rm squeezed\: probe}\\ & & \\ 
2 x^{-1}
 \rightarrow 0 & & x \gg 1 & {\rm anti-squeezed\: probe}
\end{array}
\right.
\label{checkG}\;.
\end{eqnarray}
Notice that $F$ and $G$ are {\em global} figures of fidelity \cite{kon}, {\em i.e.} compare 
the input and the output on the basis of the whole quantum state or probability
distributions rather than by their first moments, as it happens by considering
customary QND parameters (see for example \cite{milburn}, for a more general 
approach in the case of two-dimensional Hilbert space see \cite{fuc}).
\par
As a matter of fact the quantity $F+G$ is not constant, and this means that by
varying the squeezing of the probe we obtain  different trade-off between information 
gain and state disturbance. An optimal choice of the probe, corresponding to 
maximum information and minimum disturbance, maximizes $F+G$. 
The maximum is achieved for $x\equiv x_{\sc m}\simeq 1.2$, corresponding 
to fidelities $F[x_{\sc m}]\simeq 86 \%$ and $G[x_{\sc m}]\simeq 91\%$.
Notice that for a chosen signal, the optimization of the QND measurement can 
be achieved by tuning the internal phase-shift of the interferometer, without the 
need of varying the squeezing of the probe. For a nearly balanced interferometer we have 
$\tan\phi\simeq 1$: in this case the optimal choice for the probe is a state slightly 
anti-squeezed with respect to the signal, {\em i.e.} $\sigma_{\sc p} \simeq 1.2\: 
\sigma_{\sc s}$. Finally, the fidelities are equal for $x\equiv x_{\sc e}\simeq 1.3$,
corresponding to $F[x_{\sc e}]=G[x_{\sc e}]\simeq 88 \%$
\par
For non Gaussian signals the behavior is similar though no simple analytical
form can be obtained for the fidelities. In this case, in order to find the 
optimal QND measurement, one should resort to numerical means \cite{fut}. 
\section{Conclusions}\label{s:outro}
In conclusions, we have suggested an interferometric scheme assisted
by squeezing and linear feedback to realize an arbitrary QND 
measurement of a field quadrature. Compared to previous proposals the 
main features of our setup can be summarized as follows: i) it involves 
only linear coupling between signal and probe, ii) only single mode 
transformations on the conditional output are needed, and iii) the whole 
class of QND measurements may be obtained with same setup either by tuning the 
internal phase-shift of the interferometer, or by varying the squeezing of the
probe.  \par
The present setup permits, in principle, to achieve both a projective and a
fully non-destructive quantum measurement of a field quadrature.  In practice,
however, the physical constraints on the maximum amount of energy that can be
impinged into the optical channels pose limitations to the precision of the
measurements. This agrees with the facts that both an exact repeatable
measurement and a perfect state preparation cannot be realized for observables
with continuous spectrum \cite{oza}. Of course, other limitations are imposed
by the imperfect photodetection and by the finite resolution of detectors
\cite{hof}. Compared to a vacuum probe, the squeezed/anti-squeezed meters
suggested in this paper provide a consistent noise reduction in the desired
fidelity figure already for moderate input probe energy. In addition, by
varying the squeezing of the probe an optimal QND measure can be achieved,
which provides the maximum information about the quadrature distribution of
the signal, while keeping the conditional output state as close as possible to
the incoming signal.  

\newpage
\end{document}